\documentstyle[11pt,amssym,aaspp4]{article}
\received{}
\accepted{}
\journalid{}{}
\articleid{}{}
\lefthead{Ulvestad et al.}
\righthead{Sub-relativistic Seyfert Jets}
\def\H2{\ion{H}{2}}

\begin{document}

\title{Sub-Relativistic Radio Jets and Parsec-Scale \\
Absorption in Two Seyfert Galaxies }

\author{J.S.~Ulvestad\altaffilmark{1}, J.M.~Wrobel\altaffilmark{1}, 
A.L.~Roy\altaffilmark{1,2}, A.S.~Wilson\altaffilmark{3}, \\
H.~Falcke\altaffilmark{2}, \& T.P.~Krichbaum\altaffilmark{2} }
\altaffiltext{1}{National Radio Astronomy Observatory,
P.O. Box O, Socorro, NM 87801; julvesta,jwrobel@nrao.edu}
\altaffiltext{2}{Max Planck Institut f\"ur Radioastronomie,
Auf dem H\"ugel 69, D-53121 Bonn, Germany; 
aroy,hfalcke,tkrichbaum@mpifr-bonn.mpg.de}
\altaffiltext{3}{Department of Astronomy, University of Maryland,
College Park, MD 20742; wilson@astro.umd.edu}

\vskip 10pt
\centerline{\it Astrophysical Journal Letters, in press}
\vskip 10pt

\begin{abstract}
The Very Long Baseline Array has been used at 15~GHz to image 
the milliarcsecond structure of the Seyfert galaxies 
Mrk~231 and Mrk~348 at two epochs separated by about 1.7~yr.  
Both galaxies contain parsec-scale double radio sources whose
components have brightness
temperatures of $10^9$--$10^{11}$~K, implying that
they are generated by synchrotron emission.  The nuclear components 
are identified by their strong variability between epochs, indicating that
the double sources represent apparently one-sided jets.
Relative component speeds are measured to be 
$\sim 0.1c$ at separations of 1.1~pc or less 
(for $H_0 = 65$~km~s$^{-1}$~Mpc$^{-1}$),
implying that parsec-scale Seyfert jets are intrinsically different 
from those in most powerful radio galaxies and quasars.  The lack of 
observed counterjets is most likely
due to free-free absorption by torus gas, with an ionized density
$n_e\gtrsim 2\times 10^5$~cm$^{-3}$ at $T\approx 8000$~K, or
$n_e\gtrsim 10^7$~cm$^{-3}$ at $T\approx 10^{6.6}$~K, 
in the inner parsec of each galaxy.  The lower 
density is consistent with values found from X-ray absorption 
measurements, while the higher temperature and
density are consistent with direct radio imaging of the NGC~1068 
torus by Gallimore et al.

\end{abstract}


\section{Introduction}

Seyfert galaxies have weak and small radio sources, with typical 
sizes $\leq 500$~pc and typical centimeter-wavelength 
 powers of $\lesssim 10^{23}$~W~Hz$^{-1}$ (e.g., \cite{ulv89}).
These sources are apparently produced by weak
jets whose axes are determined by the galaxies' central obscuring 
disks, and interacting with thermal gas within ionization cones
(e.g., \cite{fal98}, and references therein).  The presence of 
gas disks or tori is supported by the obscuration of broad
lines in a number of Seyfert 2 galaxies 
(\cite{ant85}; \cite{mil90}; \cite{tra95}), by
VLBI imaging of H$_2$O masers in NGC~4258 
(\cite{her97}), and by parsec-scale HI absorption in 
other active galaxies (\cite{pec98}; \cite{tay99}).  

It is of considerable
interest to measure the speeds of Seyfert radio jets close to their
central engines, to attempt to differentiate between ``intrinsic'' and 
``environmental'' effects.  Previous measurements 
of jet speeds in Seyfert cores are rare. In NGC~4151, upper
limits of $0.14c$ and $0.25c$ have been measured on scales of 7
and 36~pc (\cite{ulv98}).  Component positions in NGC~1068, 
measured with the Very Large Array in 1983
(\cite{ulv87}) and with the Very Long Baseline Array (VLBA) in 1996
(\cite{roy98}), imply an upper limit of $\sim 0.5c$ for 
components $\sim 20$~pc apart.  Recently, in III~Zw~2,
Falcke et al. (1999) inferred an apparent speed
$\lesssim 0.2c$ on a sub-parsec scale, based on a single VLBI image 
following a flux outburst.

The galaxies Mrk~231 and Mrk~348 contain two of the strongest
radio sources found in Seyferts, and are prime candidates for
the measurement of motions in their cores.  Within 
larger-scale VLBI structures, each galaxy contains a parsec-scale 
double radio source (\cite{hal97}; 
Ulvestad, Wrobel, \& Carilli 1999).
Those central sources have now been imaged
at two epochs using the VLBA;\footnote{The VLBA is part 
of the National Radio Astronomy Observatory, a facility of the National
Science Foundation operated under cooperative agreement
by Associated Universities, Inc.} 
this {\it Letter} reports measurements of the
component separation speeds on scales of $\leq 1.1$~pc.

\section{VLBA Observations \& Data Analysis}
\label{vlbaobs}

The VLBA (\cite{nap94}) was used to observe
Mrk~231 and Mrk~348 at two epochs separated by $\sim 1.7$~yr, between 
1996 and 1998.  For each galaxy, observing frequencies ranged from 1.4 to 
22~GHz; the images at 15.365~GHz, obtained in
left-circular polarization, have the best combination of resolution and
sensitivity for the compact components, and are the subject
of this {\it Letter}.  
All data were initially calibrated in AIPS (\cite{van96}), then 
iteratively imaged and self-calibrated in DIFMAP (\cite{she97}).
On-source integration times ranged from 1.2 to 2.6~hr, resulting
in r.m.s. noises of 0.4--1.1~mJy~beam$^{-1}$ for the final
uniformly weighted images.

Two-component Gaussian models
were fitted in both the ({\it u,v}\/) plane and the image plane;
we use the 
image-plane fits here.  Flux-density errors (all errors are $1\sigma$)
were derived by combining a 5\% scale uncertainty in
quadrature with the fitting error (including noise).  
Estimated size errors are 20\% in each axis, 
with point sources taken to have upper limits of half the beam size.  
Errors in relative component positions are consistent with 
the measured noises, except for the first-epoch image of Mrk~231, 
where the quoted error is the total range found by using different 
fitting procedures. 

\section{Two-Epoch Radio Images}

\subsection{Mrk 348}

\begin{figure}[htp]
\vspace{14cm}
\includegraphics{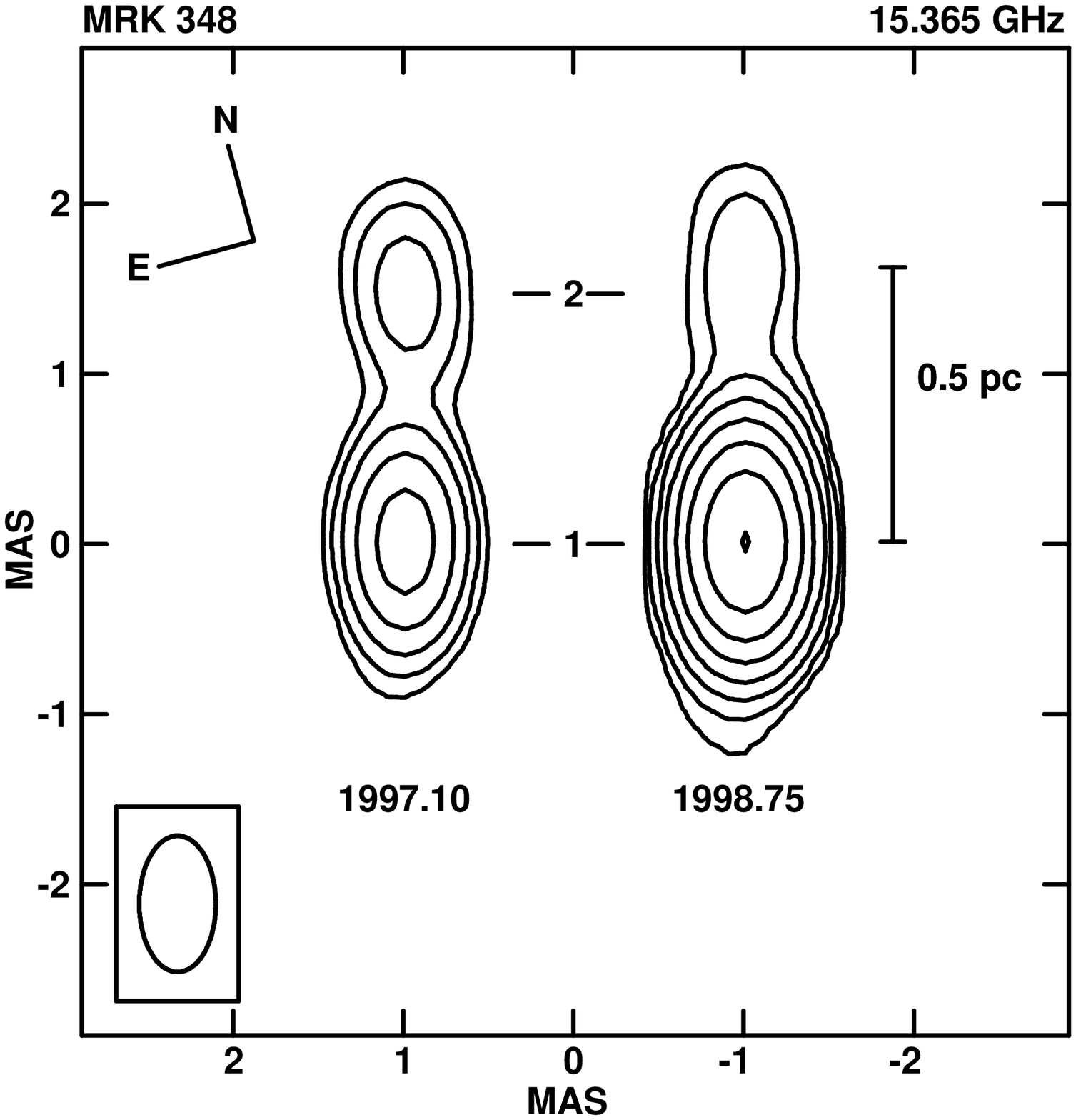}
\caption{VLBA 15-GHz images of Mrk 348 at epochs 1997.10 and 1998.75.  The
images have been rotated by 15\arcdeg\ from the cardinal orientation, 
aligned at the southern radio component,
then offset from each other horizontally by 2~mas. North and East
are labeled, as are the two radio components.  Logarithmic contours
start at 4~mJy~beam$^{-1}$ and increase
by factors of 2 to 512~mJy~beam$^{-1}$.  The common restoring beam is
is $0.80\times 0.45$~mas in PA $-15^\circ$.}
\label{fig:348-im}
\end{figure}

\placefigure{fig:348-im}
\placetable{tab:prop}

Mrk 348 has a redshift $z=0.014$ (\cite{dev91}), yielding a scale of 
0.31~pc~mas$^{-1}$ for $H_0=65$~km~s$^{-1}$~Mpc$^{-1}$ (used throughout).
This galaxy is a type 2 Seyfert with a hidden broad-line
region (\cite{mil90}), implying that its inner disk is 
nearly edge-on. It also contains a 200-mas triple
radio source (\cite{nef83}) coinciding with optical 
[O~III] emission imaged by Capetti et al. (1996).
Our VLBA images (Figure~\ref{fig:348-im}) show a small-scale double source 
that is aligned with the larger scale radio and optical emission.
Phase referencing at the second epoch gives a J2000 position for
the stronger component of $\alpha = 00^h48^m47^s$\llap{.}1422, 
$\delta = 31^\circ 57'25''$\llap{.}044, with
an error of $\sim 12$~mas.
The relative component separation (see Table~\ref{tab:prop}) increased
from 1.46~mas to 1.58~mas in 1.65~yr.
Since the stronger component was resolved at the second epoch, 
its centroid could have shifted in absolute position; we estimate
a $3\sigma$ upper limit 
to this shift that is equal to the component size of 0.16~mas.
Adding this error in quadrature to the nominal position
errors, we find a proper 
motion of $0.073\pm 0.035$~mas~yr$^{-1}$, for an 
apparent speed (in units of $c$) of $\beta_{\rm app} = 0.074\pm 0.035$.
The implied epoch of zero separation is
$1977^{+7}_{-20}$, so the secondary might have had its genesis during a 
strong flux outburst in early 1982 (\cite{nef83}).  

The total flux density of $122\pm 6$~mJy at 1997.10 was slightly smaller 
than the flux density of $169\pm 9$~mJy measured at 1995.26 
(\cite{bar98}). Mrk~348 since has undergone a major radio flare,
with the southern component increasing by a factor of 5.5 between 1997.10
and 1998.75, strongly suggesting that it is the galaxy nucleus.  This is 
similar to the more extreme flares in III~Zw~2, which 
are discussed by Falcke et al. (1999).

\subsection{Mrk 231}

\begin{figure}[htp]
\vspace{14cm}
\includegraphics{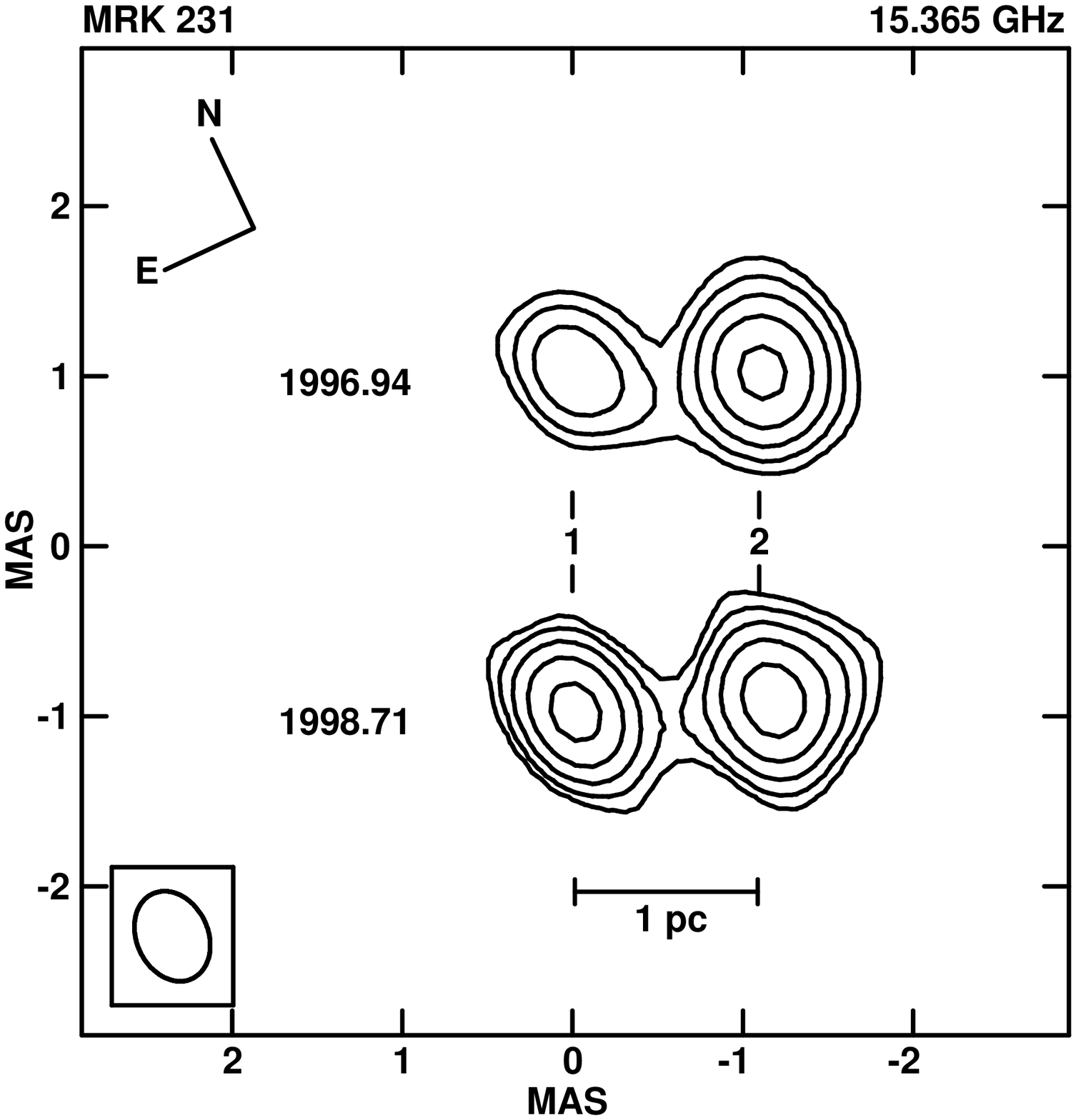}
\caption{VLBA 15-GHz images of Mrk 231 at epochs 1996.94 and 1998.71.  The
images have been rotated by 25\arcdeg\ from the cardinal orientation, 
aligned at the northeastern radio component,
then offset from each other vertically by 2~mas.  North and East
are labeled, as are the two radio components.  Logarithmic contours
start at 2~mJy~beam$^{-1}$ and increase
by factors of 2 to 32~mJy~beam$^{-1}$.  The common restoring beam 
is $0.55\times 0.42$~mas in PA $0^\circ$.}
\label{fig:231-im}
\end{figure}

\placefigure{fig:231-im}

The redshift of Mrk~231 is $z=0.042$ (\cite{dev91}), and the corresponding 
scale is 0.93~pc~mas$^{-1}$.  Mrk~231 is a Seyfert 1/starburst galaxy
with a heavily obscured nucleus and a total infrared luminosity 
of $\sim 3\times 10^{12}L_\odot$ (\cite{soi89}).  It
contains a 40-pc north-south radio source (\cite{nef88};
\cite{ulv99}) embedded within a starburst several hundred
parsecs in extent (\cite{bry96}; \cite{car98}).  Our VLBA images
(Figure~\ref{fig:231-im})
show a double source with a position angle differing by about 
65\arcdeg\ from the larger scale source.  The second-epoch 
phase-referenced J2000 position ($\alpha = 12^h56^m14^s$\llap{.}2336,
$\delta = 56^\circ 52'25''$\llap{.}245)
is consistent within the 12-mas error 
with that previously listed by Patnaik et al. (1992).
The increase in component separation (Table~\ref{tab:prop})
results in a measured proper motion of $0.046\pm 0.017$~mas~yr$^{-1}$,
or $\beta_{\rm app}=0.14\pm 0.052$, with a zero-separation epoch
of $1973^{+7}_{-15}$.  Variability by a factor of 2.5
between epochs indicates that the weaker, northeastern component
is the actual nucleus of the galaxy.  

\section{Nature of the One-Sided Sources}

Parsec-scale radio sources have now been imaged in several Seyfert 
galaxies (e.g., \cite{gal97}; \cite{ulv98}).
In NGC~1068, the parsec-scale source is believed to represent thermal
emission from the accretion torus (Gallimore et al. 1997). However, 
the brightness temperatures in Mrk~348 and Mrk~231 are too high for
thermal emission, and instead suggest association
of the radio components with outflowing jets.  Strong variability
in one component in each galaxy indicates that it
is close to the active nucleus, and that
the double sources represent one-sided jets rather than 
straddling the galaxy nuclei.  

\subsection{Relativistic Boosting?}

One-sided radio structures often are caused by relativistic jets
having speed $\beta c$ and pointing at a small angle $\theta$ with 
respect to the observer's line of sight (for equations,
see \cite{pea87}).  We assume an intrinsic spectral index of
$\alpha=-0.7$ ($S_\nu\propto\nu^{+\alpha}$) for the off-nuclear
components.  In Mrk 348, therefore, the observed jet/counterjet ratio of
$R>17$, together with $\beta_{\rm app}\approx 0.08$, can be due to relativistic
boosting only if $\beta\gtrsim 0.37$ and $\theta \lesssim 10^\circ$.  
However, the inner disk is edge-on (\cite{mil90}), and the half-angle of the ionization
cone is $\sim$45\arcdeg\ (\cite{sim96}); if the radio jet is inside that 
cone, $\theta \gtrsim 45^\circ$, inconsistent with motion near the line of sight.
In Mrk~231, $R>45$ and $\beta_{\rm app}\approx 0.14$, requiring 
$\beta\gtrsim 0.48$ and $\theta \lesssim 10^\circ$ for Doppler boosting to
account for the one-sided source.  
If the radio axis is perpendicular to the 100-pc-scale disk 
(\cite{bry96}; Carilli et al.~1998), then $\theta\approx 45^\circ$, 
also inconsistent with motion at a very small viewing angle.

The above arguments depend on two assumptions: (1) the jets flow along
the axes of the inner disks, as they apparently do in
NGC~1068 (Gallimore et al. 1997) and NGC~4258 (\cite{her97});  
and (2) the measured speeds represent
the actual jet speeds, rather than quasi-stationary structures
(such as shocks) through which faster jets flow.
The interpretation of fast jets moving through slower shocks has 
been made for Centaurus~A, with $\beta_{\rm app}\sim 0.1$, 
based on the internal evolution of radio components (\cite{tin98}).  
However, in Mrk~231 and Mrk~348, the structures of the off-nuclear
components are consistent between the two epochs, so there is no similar 
evidence for higher flow speeds.

\subsection{Free-Free Absorption}

A straightforward explanation for the one-sided sources is that the 
``missing'' components are in the receding jet and are free-free 
absorbed by ionized gas.  The optical depth at frequency $\nu$ 
is 
\begin{equation}
\tau_{\rm ff}(\nu)\ \approx\ 8.235\times 10^{-2}\ T^{-1.35}\ 
(\nu/{\rm GHz})^{-2.1}\ E\
\end{equation}
(\cite{mez67}), where $T$ is the temperature in Kelvin and
$E$ is the emission measure in cm$^{-6}$~pc.  
Optical depths of $\tau_{\rm ff}(15\ {\rm GHz}) \gtrsim 4$ are needed
in order 
to account for the observed jet/counterjet ratios.  Assuming a line-of-sight 
distance of $\sim 0.1$~pc through the ionized gas, and a gas
temperature of 8000~K, the average ionized densities would
be $n_e \gtrsim 2\times 10^5$~cm$^{-3}$ at 0.5--1~pc from the 
galaxy nuclei.  The resulting
column densities of $\sim 10^{23}$~cm$^{-2}$ are remarkably
consistent with the measured X-ray absorption columns
in Mrk~348 (\cite{smi96}) and in Mrk~231 (\cite{nak98}).  The
density also is consistent with that found for possible absorption
in a warm, weakly ionized medium in the H$_2$O-maser galaxy 
NGC~2639 (\cite{wil98}).
On the other hand, if the absorption comes from
a much hotter gas having $T\approx 10^{6.6}$~K,
the average density required would be $n_e \gtrsim 10^7$~cm$^{-3}$.
These values are close to those
inferred from the image of the torus radio emission
in NGC~1068 (Gallimore et al. 1997); such disk emission 
in Mrk~231 and Mrk~348 is possible, since it would not 
be detectable by our observations, due to inadequate resolution
and brightness-temperature sensitivity.

The free-free-absorption interpretation requires ionized gas 
densities of $10^5$--$10^7$~cm$^{-3}$
in the inner parsec of Mrk~231 and Mrk~348.  These are  
higher than the values of $n_e\sim 10^4$~cm$^{-3}$ inferred
from the absorption $\sim 2$~pc from the nucleus of 3C~84 (\cite{lev95}), 
and $n_e \sim 10^3$~cm$^{-3}$ inferred for absorption 15--20~pc from
the nucleus of Mrk~231 (\cite{ulv99}).
Thus, our results are consistent with the presence of
disks or tori having ionized densities that fall gradually from 
$10^5$--$10^7$~cm$^{-3}$ in the inner parsec of the galaxies to 
$\sim 10^{3}$~cm$^{-3}$ at $\sim 20$~pc from the nuclei.
Even though no H$_2$O maser emission has been detected in Mrk~231 
or Mrk~348 (\cite{bra96}), possibly because of our viewing angle, 
the one-sided sub-relativistic jets are consistent with the presence 
of megamaser disks.  If $\tau_{\rm ff}(15\ {\rm GHz})\approx 4$, 
the absorbing gas may become optically thin near 30~GHz, and 
counterjets might be detectable in very sensitive VLBI observations 
at 43~GHz.

\subsection{Comparison to Other Parsec-Scale Sources}

The apparently low jet speeds in Mrk~231 and Mrk~348 are similar to 
those seen on parsec scales in some weak Fanaroff-Riley I 
(\cite{fan74}) radio galaxies such as Centaurus~A (\cite{tin98}) and 3C~84 
(\cite{dha98}).  However, most such objects are relativistic 
on parsec scales (\cite{gio98}), as are most strong radio galaxies and quasars 
(\cite{pea96}).  The apparently sub-relativistic Seyfert jet speeds on the
same scale may be due to interactions with the small-scale gas, or to
physics directly related to the energy source, such as
low black-hole spin rates (e.g., \cite{ree82}; \cite{wil95}).  

Both Mrk~348 and Mrk~231, as well as NGC~2639 (\cite{wil98})
and III~Zw~2 (\cite{fal99}), have radio spectra that peak at 
10~GHz or higher.  Thus,
the Seyfert radio sources have general characteristics similar to
compact symmetric objects (CSOs), which display gigahertz-peaked
spectra, symmetric radio sources on
scales of $\sim 100$~pc, and one-sided jets on scales of 10~pc
(\cite{tay96}).  Small-scale Seyfert jet speeds also are similar to those 
measured in CSOs at 50--100~pc from their nuclei 
(\cite{ows98a}; \cite{ows98b}).  The CSO jets are one-sided at 15~GHz
(Taylor et al.~1996), and even at 43~GHz (Taylor, private communication),
on scales of 10~pc; at least one CSO also shows patchy H{\sc I} 
absorption on similar scales (\cite{pec99}).  We speculate that
free-free absorption with column densities near $10^{24}$~cm$^{-2}$ 
might cause one-sided structures
in some CSOs, if their disks are larger and denser than in Seyferts.
High-energy X-ray spectral studies could be used
to search for absorption due to this gas.

\acknowledgments

We thank Greg Taylor, Alison Peck, and Jack Gallimore for useful
discussions and suggestions, and Daria Halkides for assistance with the
data reduction for the first epoch of Mrk~348.  H.~Falcke acknowledges
support from DFG grants Fa~358/1-1 and 1-2.
This research has made use of the NASA/IPAC
Extragalactic Database (NED) which is operated by the Jet Propulsion
Laboratory, California Institute of Technology, under contract with
the National Aeronautics and Space Administration.

\clearpage

\begin{deluxetable}{cccccc}
\tablecolumns{6}
\tablewidth{0pc}
\tablecaption{Results of 2-Component Gaussian Fits}
\tablehead{
\colhead{No.}&\colhead{Flux Density}&\colhead{Size}&\colhead{$T_b$\tablenotemark{a}}&
\colhead{Offset}&\colhead{PA} \\
&\colhead{(mJy)}&\colhead{(mas$\times$mas, deg)}&\colhead{($10^9$ K)}&
\colhead{(mas)}&\colhead{(deg)} }
\startdata
\multicolumn{6}{c}{\underbar{Mrk 348, 1997.10}} \\
1 (S)&$96\pm 5$&Unresolved&$>6.7\pm 1.9$&0.000&\nodata \\
2 (N)&$26\pm 2$&Unresolved&$>1.8\pm 0.5$&$1.460\pm 0.009$&$-16\pm 1$ \\
\multicolumn{6}{c}{\underbar{Mrk 348, 1998.75}} \\
1 (S)&$552\pm 28$&$0.16\times 0.11$, PA $-9\pm 4$&$238\pm 68$&0.000&\nodata \\
2 (N)&$17\pm 1$&Unresolved&$>1.8\pm 0.5$&$1.581\pm 0.021$&$-15\pm 2$ \\
\multicolumn{6}{c}{\underbar{Mrk 231, 1996.94}} \\
1 (NE)&$17\pm 1$&Unresolved&$>2.2\pm 0.6$&0.000&\nodata \\
2 (SW)&$51\pm 3$&$0.32\times 0.18$, PA $-74\pm 4$&$6.9\pm 2.0$&$1.081\pm 0.030$&
$-115\pm 1$ \\ 
\multicolumn{6}{c}{\underbar{Mrk 231, 1998.71}} \\
1 (NE)&$44\pm 3$&Unresolved&$>6.0\pm 1.7$&0.000&\nodata \\
2 (SW)&$60\pm 3$&$0.30\times 0.13$, PA $-88\pm 4$&$12.0\pm 3.4$&$1.162\pm 0.004$&
$-112\pm 1$ \\
\enddata
\tablenotetext{a}{All brightness temperatures are expressed in
the source rest frames, by multiplying the observed brightness
temperatures by $(1+z)$.}
\label{tab:prop}
\end{deluxetable}
%\clearpage

\clearpage

\clearpage

\end{document}